\documentclass{article}

\usepackage{url}
\usepackage{xr}
\makeatletter
\newcommand*{\addFileDependency}[1]{
  \typeout{(#1)}
  \@addtofilelist{#1}
  \IfFileExists{#1}{}{\typeout{No file #1.}}
}
\makeatother

\usepackage{arxiv}

\usepackage[utf8]{inputenc} 
\usepackage[T1]{fontenc}    
\usepackage{hyperref}       
\usepackage{url}            
\usepackage{booktabs}       
\usepackage{amsfonts}       
\usepackage{nicefrac}       
\usepackage{microtype}      
\usepackage{lipsum}		
\usepackage{graphicx}
\usepackage{natbib}
\usepackage{doi}

\usepackage{fancybox}

\usepackage{bm}
\usepackage{lscape} 
\usepackage{multirow}
\usepackage{pifont}
\usepackage{schemata}
\usepackage{booktabs}
\usepackage{xcolor} 
\usepackage{bbm}
\usepackage{arydshln} 
\usepackage{etoolbox} 

\usepackage{mathtools}

\AtBeginEnvironment{tablenotes}{\smallskip\footnotesize}
\newcommand{\pkg}[1]{\texttt{#1}}
\newcommand{\bx}{\bm{x}}
\newcommand{\bbeta}{\bm{\beta}}
\newcommand{\CIF}{\text{CIF}}

\DeclareMathOperator*{\argmin}{argmin}

\newcommand{\Exp}{\text{Exp}}
\newcommand{\Gammad}{\text{Gamma}}
\newcommand{\Norm}{\text{N}}
\newcommand{\GG}{\text{G}\text{Gamma}}

\DeclareMathOperator{\E}{\mathbb{E}}

\usepackage{threeparttable}

\title{A review on competing risks methods for survival analysis}

\author{ {\hspace{1mm}Karla Monterrubio-G\'omez}\thanks{Corresponding author} \\
	MRC Human Genetics Unit\\
	University of Edinburgh\\
	Edinburgh, United Kingdom\\
	\texttt{kmonterr@ed.ac.uk} \\
	\And
	{\hspace{1mm} Nathan Constantine-Cooke} \\
	MRC Human Genetics Unit\\
	University of Edinburgh\\
	Edinburgh, United Kingdom\\
	\texttt{nathan.constantine-cooke@ed.ac.uk} \\
	 \AND
	{\hspace{1mm} Catalina A. Vallejos} \\
	MRC Human Genetics Unit   \hspace{9mm}  The Alan Turing Institute\\
	{\color{white} edn}University of Edinburgh \hspace{8mm}  {\color{white} e}London, United Kingdom\\
	Edinburgh, United Kingdom \hspace{9mm} {\color{white} London, United Kingdom}\\
	\texttt{catalina.vallejos@ed.ac.uk} \\
}

\renewcommand{\shorttitle}{A review on competing risks methods for survival analysis}

\hypersetup{
pdftitle={A review on competing risks methods for survival analysis},
pdfsubject={q-bio.NC, q-bio.QM},
pdfauthor={Karla Monterrubio-G\`omez, Nathan Constantine-Cooke, Catalina A. Vallejos},
pdfkeywords={First keyword, Second keyword, More},
}

\begin{document}
\maketitle

\begin{abstract}
When modelling competing risks survival data, several techniques have been proposed in both the statistical and machine learning literature. State-of-the-art methods have extended classical approaches with more flexible assumptions that can improve predictive performance, allow high dimensional data and missing values, among others. 
Despite this, modern approaches have not been widely employed in applied settings.
This article aims to aid the uptake of
such methods by
providing a condensed compendium of competing risks survival methods with a unified notation and interpretation across approaches. We highlight available software and, when possible, demonstrate their
usage via reproducible R vignettes. 
Moreover, we discuss two major concerns that can affect benchmark studies in this context: the choice of performance metrics and reproducibility.
\end{abstract}


\section{Introduction}
\label{S:1}
Survival or time-to-event analysis comprises a collection of methods to model the time until an event of interest occurs. Usually, the goal is to estimate the risk of observing the event by a given time or to quantify the relationship between event risk and known covariates. 
Survival methods 
are widely used in several fields; including medicine, social sciences, engineering and economics. Reviews about such methods are provided by \cite{cox1984analysis}, \cite{Carpenter1997},  \cite{klein2006survival} and, more recently, \cite{Wang2019}. 

A typical element of survival data is \emph{censoring}, where event times are unknown.  
This can occur for several reasons, e.g.~lost of follow-up. 
Survival methods such as the popular 
Cox proportional hazards (CPH) model \citep{Cox1972} often assume independent censoring: those who were censored at a specific time are representative of all those who remained at risk.


In some cases, a subject can experience more than one type of mutually exclusive events --- typically referred to as \emph{competing risks} (CR). For instance, 
a patient can die from different causes (e.g. cancer or non-cancer death). If the main focus is a specific event type, others could be recorded as censored observations. 
However, the independent censoring assumption does not hold in this setting: if one event occurs, the others are no longer possible. This can lead to biased estimates in standard survival models \citep{Austin2016}. 

The development of CR survival models is an active area of research \citep[e.g.][]{ng2003based, Ishwaran2014, lee2018deephit, Siames2018, Dauda2019, Bart2020}, but state-of-the-art approaches have not been widely adopted in applied settings. 
This may be because papers are not aimed for practitioners, or due to lack of clear benchmarks that highlight the strengths and drawbacks of each method. The lack of (open-source) software can also prevent wide adoption.
As a result, real-world applications of CR survival models have primarily made use of long-established methods \citep[e.g.][]{FineGray1999}, leaving the application of cutting-edge methodologies often limited only to academic exercises.

The purpose of this review is to summarise the current landscape of 
CR approaches, 
including 
methods  
developed by two overlapping but still distinctive communities; namely, statistics and machine learning. We aim to unify the notation and interpretation across methods, facilitating their comparison. To aid the uptake of state-of-the-art tools, we highlight available software and, when possible, demonstrate their use via reproducible \pkg{R} vignettes (see \url{www.github.com/KarlaMonterrubioG/CompRisksVignettes}). We also discuss common issues encountered when evaluating new CR methods; such as reproducibility and the choice of performance metrics.


\section{Background}
\label{S:background}

Consider a continuous random variable, $T \geq 0$, defined as the time until which an event of interest occurs. Let $f(t)$ be the probability density function for $T$.  
Often, survival models are specified via the survival function $S(t) = \Pr(T>t) = \int_t^{\infty} f(t) \,d t$ or the hazard function \begin{equation} \label{eq:hazard} h(t) = \lim_{\Delta t\to 0} \frac{ \Pr( t<T \leq t + \Delta t \mid T>t)}{ \Delta t} = \frac{f(t)}{S(t)},\end{equation}
i.e. the instantaneous rate, 
given that no event has occurred by time $t$. 
A variety of parametric and non-parametric methods exist when a single event type can occur. In the presence of multiple event types (e.g.~cancer/non-cancer death), a composite event (e.g.~all cause mortality) can be defined at the cost of reduced data granularity. Instead, CR survival models can explicitly capture different event types. Here, we focus on 
mutually exclusive events: any event prevents the others. If one event prevents others but not vice-versa (e.g. myocardial infarction and death), a semi-CR setting may be used \citep{fine2001semi, SEMI2007, semi2008}. 



\subsection{Competing risks survival models}


Assume $K$ event types and let $Z \in \{0, \ldots, K\}$ be a random variable representing the observed type of event (typically $Z = 0$ denotes censoring).
Different frameworks have been used to define CR survival models. First, using the cause-specific (CS) hazard function, which quantifies the instantaneous rate for the $k$-th event type for subjects that have not experienced \emph{any} event: \begin{equation}
    h_k^{\text{{CS}}}(t)= \lim_{\Delta t\to 0} \frac{ \Pr( t<T \leq t + \Delta t, Z = k \mid T>t)}{ \Delta t}.
 \label{eq:H_CS}
\end{equation} 

The overall hazard in~\eqref{eq:hazard} is the sum across all CS hazards, i.e.~$h(t) =\sum_{k=1}^{K} h_k^{\text{{CS}}}(t)$. Alternatively, CR survival models can also be defined via 
the cumulative incidence function (CIF): 
\begin{equation} \label{eq:CIF}
\CIF_k(t)=\Pr( T \leq t , Z = k),
\end{equation} i.e. the probability of observing the $k$-th event type before time $t$ (and prior to other events) or 
the sub-distribution hazard function \citep[often referred to as the Fine-Gray hazard,][]{gray1988}
\begin{equation}
    h_k^{\text{{FG}}}(t) = - \frac{d \log (1- \CIF_k(t))}{dt} = \lim_{\Delta t\to 0} \frac{ \Pr( t<T \leq t + \Delta t, Z =  k \mid T>t \cup (T < t \cap Z \neq k))}{ \Delta t}, 
    \label{eq:H_FG}
\end{equation}
quantifying the instant rate of the $k$-th event for subjects that have not had \emph{that} event by time $t$, but including those who experienced a competing event. Note that, the key difference between \eqref{eq:H_CS} and  \eqref{eq:H_FG} is the risk set used to define the probability. 

 Finally, 
 latent failure times CR models assume 
 $T= \min\{T_1, \ldots, T_K \}$ and $Z= \argmin_k \{ T_k\}$, where $T_k$ is an event-specific time which is unobserved, unless $Z=k$.  
Such models are typically defined through the joint survival function 
$S_{T_1,\ldots,T_K} (t_1,\ldots,t_K)= \Pr(T_1>t_1,\ldots T_K>t_K)$. However, the marginal distributions of the latent times $T_k$ are non-identifiable, unless non testable assumptions (e.g.~independence between $T_k$'s \citep{ cox1962renewal, Tsiatis} or that dependency arises through a known copula \citep{copula}) are made. 

\subsection{Regression models for CR survival data} \label{subsec:regressionCR}
Often the aim is to quantify how a set of covariates (features) affects CR outcomes, or to use such covariates in order to predict the risk associated to different event types. Assume we have observations $\{(T_i, Z_i), i = 1, \ldots, n\}$, where $T_i$ and $Z_i$ represent the event time and event type for the $i$-th subject. 
Let $\bx_i \in \mathbb{R}^{p}$ be a $p$-dimensional vector of features for subject $i$. As in \cite{Cox1972}, a regression model can be defined via the CS hazard functions in \eqref{eq:H_CS} as 
\begin{equation}\label{eq:CPH}
      h_k^{\text{{CS}}}(t_i \mid \bx_i) = h_{k0}^{\text{{CS}}}(t_i) \exp \left(   \bx^\top_{i} \bbeta_{k} \right), \quad k=1,\ldots K,
\end{equation} 
where $h_{k0}^{\text{{CS}}}(\cdot)$ is a CS baseline hazard 
and $\bbeta_k=(\beta_{k1}, \ldots \beta_{kp})^\top$ a vector of covariate effects such that $\exp\left( \beta_{kj}\right)$ is the relative change in the CS hazard linked to a unit change in the $j$-th covariate. Inference can be done by fitting $K$ separate CPH regressions where competing events are treated as censored observations. An estimate of $h_{k0}^{\text{{CS}}}(\cdot)$ is not required to infer $\bbeta_k$ \citep{Cox1972}, 
but it is needed to perform prediction and an estimator in \cite{breslow1972discussion} can be used for this purpose. 

\cite{FineGray1999} developed an alternative approach based on the sub-distribution hazard function in \eqref{eq:H_FG}. 
Analogous to \eqref{eq:CPH}, this is defined by
\begin{equation}\label{eq:FG}
 h_k^{\text{{FG}}}(t_i \mid \bx_i) = h_{k0}^{\text{{FG}}}(t_i) \exp \left(  \bx^\top_{i} \bm{\gamma}_k \right), \quad k=1,\ldots K,
 \end{equation} 
 where $\bm{\gamma}_k =(\gamma_{k1}, \ldots, \gamma_{kp})^\top$ is a vector of covariate effects estimated using the inverse probability weighting \citep{Robins1992}. These are not the same as those in \eqref{eq:CPH}; thus, one should be cautious with their interpretation \citep{Austin2017}. The sign of $\gamma_{kj}$ indicates whether an increase in the $j$-th covariate is associated with an increase/decrease in the incidence of the event, but $\gamma_{kj}$ does not measure effect sizes on the probability of the occurrence of the event. 
 Due to \eqref{eq:H_FG}, \eqref{eq:FG} is often referred as a CIF regression model and can be re-written as: 
 \begin{equation}
 \log \left[ - \log\left[1 - \CIF_k(t_i \mid \bx_i) \right] \right] = \log \left[ - \log\left[ 1 - \CIF_{k0}
 (t_i) \right] \right] + \bx^\top_{i} \bm{\gamma}_k, \hspace{0.5cm} k = 1, \ldots, K, \label{eq:CIF_reg}
 \end{equation} where $\CIF_{k0}(\cdot)$ is the baseline CIF for the $k$-th event (all covariate values equal to zero). This can be interpreted as a Generalized Linear Model (GLM) with a complementary log‐log link function. The CIF regression approach is better suited than \eqref{eq:CPH} when developing risk prediction models \citep{Austin2016}. However, one limitation is that, for certain covariate and time specifications, the sum of the $K$ estimated CIFs may exceed 1 \citep{Austin2021}.

 
Regression models based on latent failure times also exist. For example, under independence, an {\it{accelerated failure time}} (AFT) model \citep{kalbfleisch2011statistical} can be used for each latent time. Let $\log(T_{ik})$ be the $k$-th latent time for subject $i$. The AFT model can be defined as  \begin{equation} \log(T_{ik}) = \bx_i^\top\bm{\nu}_k+\varepsilon_{ik}, \end{equation} 
where 
 $\bm{\nu}_k = (\nu_{k1}, \ldots \nu_{kp})^\top$ is a vector of regression parameters 
 and $\varepsilon_{ik}$ are independent and identically distributed errors. 
 Depending on the error distribution, 
 several parametric models can be obtained (e.g.~Weibull or log-Normal). In addition, in the case of dependent latent failure times, \cite{Heckman1989} provide identifiability conditions for both PH and AFT models.

\section{Recent advances on competing risks survival models} \label{S:CR}

Recent CR methods have introduced flexibility in terms of non-linear covariate effects, time varying covariates, variable selection, missing data, and scalability, among others. Here, we summarise state-of-the-art approaches for CRs. Previous reviews in this area \citep[e.g.~][]{zhang2008modeling,haller2013} have primarily focused on the statistics literature. Instead, we provide a more comprehensive survey which covers recent contributions made by the machine learning community. As the boundary between these disciplines is diffuse \citep{bzdok2018}, we do not explicitly distinguish them. Instead, methods are grouped based on the specifications discussed in Section \ref{subsec:regressionCR} (see Table \ref{tab:ML_summary} for a summary of the methods included in this review).

\subsection{Approaches based on a proportional cause-specific hazard specification}

\subsubsection{Sparse regression.}\label{subsub:CS_sparse}
As mentioned in Section \ref{subsec:regressionCR}, the model in \eqref{eq:CPH} can be estimated 
using available software for CPH models. If $p<n$, but large with respect to $n$, 
this could lead to 
overfitting. Moreover, this is not possible in high-dimensional settings ($p>n$). 
Penalisation strategies can alleviate these problems by  
shrinking 
regression
coefficients towards zero. This is achieved by adding a penalty term to 
the partial likelihood 
that is used to fit CPH models. 
Several penalised regression methods have been proposed, e.g.~ {\it{lasso}} \citep{Tibshrani97}, {\it{adaptive lasso}} \citep{zhang2007}, {\it{elastic net}}
\citep{Engler2009} and {\it{scad}}
\citep{fan2002}.

Alternatively, one can use an ensemble learning 
strategy such as Cox model-based boosting \citep{ridgeway1999} or Cox likelihood-based boosting \citep{binder2008}. 
These methods introduce sparsity by permitting the inclusion of mandatory and optional covariates. The first approach incorporates mandatory features through an offset term \citep{boulesteix2010testing}, where the regression coefficients for the mandatory covariates are not updated during the boosting procedure. In contrast, the method by \cite{binder2008} permits updating the regression coefficients of both, mandatory and optional covariates, but the optional features may be excluded through penalisation. \citet[][Section 4.2 and 4.3]{Riccardo2016} pointed out how these different strategies can be reformulated in an equivalent manner.

\subsubsection{Lunn-McNeil.}
Instead of modelling event types separately, \cite{lunn1995} propose 
a joint model. To do so, the data is converted into an augmented layout which is
constructed as follows. 
For individuals that experienced one of the $K$ event types, the data is duplicated $K-1$ times, setting the event as censored for the repeated observations.
For censored observations, $K$ rows are added with the event 
marked as censored. 
See Appendix~\ref{Sup:C} for an example. 

Two frameworks for inference are introduced. First, a \emph{stratified} approach, which is equivalent to separate CPH models for each event type and where the $k$-th hazard is given by \begin{equation}
 h_k^{\text{LM1}}(t_i \mid \bx_i) = h_{k0}^{\text{LM1}}(t_i) \exp \left( \sum_{k=1}^K \delta_{ik}\bx_i^\top\bm{\beta}_k \right).  \label{eq:LM1}  
\end{equation}

In \eqref{eq:LM1}, $\delta_{ik} = \mathbbm{1}\{ Z_i = k \}$ are event type indicators, $h_{k0}^{\text{LM1}}(\cdot)$ is a CS baseline hazard and $\bm{\beta}_k$ denotes a $p$-dimensional vector of regression coefficients. Unlike cases in which separate models are fit for each event type, this approach permits to explore simpler models, e.g. where $\bm{\beta}_1 = \bm{\beta}_2$. 

The second, 
\emph{unstratified}, 
framework is defined as 
\begin{equation}
   h_k^{\text{{LM2}}}(t_i \mid \bx_i) = h_{0}^{\text{{LM2}}}(t_i) \exp \left(\bx^\top_{i} \bm{\theta}^\prime + \sum_{k=2}^K  \alpha_k\delta_{ik} +   \sum_{k=2}^K \delta_{ik}\bx^\top_{i} \bm{\beta}^\prime_k \right),  \label{eq:LM2} 
\end{equation}
 where the overall baseline hazard $h_{0}^{\text{{LM2}}}(\cdot)$ and the $p$-dimensional vector of coefficients $\bm{\theta}^\prime$ relate to the first event type ($k=1$), which is used as a reference. In \eqref{eq:LM2}, $\alpha_k$ and $\bm{\beta}^\prime_k$ $ (k=2,\ldots,K)$ capture deviations (baseline hazards and covariate effects) with respect to the reference event. 
 
 


\subsection{Approaches based on the CIF}

\subsubsection{Sparse regression.}
Similar to the methods described in Section \ref{subsub:CS_sparse}, approaches that adapt \eqref{eq:CIF_reg} to high-dimensional scenarios ($p>n$) have been proposed. For instance, \cite{fu2017penalized} introduced a general penalised regression framework that permits individual and grouped variable selection. The authors studied several types of penalties, such as
{\it{lasso}} \citep{Tibshrani97}, {\it{adaptive lasso}} \citep{zhang2007}, {\it{scad}} \citep{fan2002} and {\it{mcp}} \citep{zhang2010}. In this setting, inference was implemented using a modification of the coordinate descent algorithm. 


Alternatively,
\citet{binder2009} 
propose a {\it{sub-distribution hazard boosting}} approach, where
covariates are divided into a set of mandatory ($\mathcal{I}^{\text{mand}}$) and a set of optional ($\mathcal{I}^{\text{opt}}$) features. 
At each boosting iteration, $b =1, \ldots B$, regression coefficients for mandatory features $\gamma_{kl} (\forall l \in \mathcal{I}^{\text{mand}})$ are estimated jointly by maximising the partial likelihood. Then, for optional covariates, only one regression parameter 
is updated.
The selection of which coefficient to update is based on penalised partial log-likelihood estimates for all possible models:
\begin{equation}
 h_k^{\text{{FG}}}(t_i \mid \bx_i) = h_{k0}^{\text{{FG}}}(t_i) \exp \left(  \zeta^{(b-1)} _{ki}+ x_{ij}\eta^{(b)}_{kj}\right), \quad \zeta^{(b-1)} _{ki}= \bx_i^\top \bm{\gamma}_k^{(b-1)}, \quad   j \in \mathcal{I}^{\text{opt}}, 
 \end{equation} 
where $\zeta^{(b-1)} _{ki}$ is treated as an offset. 
Regression coefficients are then updated as 
$\gamma^{(b)}_{kj} = \gamma^{(b-1)}_{kj}+ \eta^{(b)}_{kj}$ for the selected covariate and $\gamma^{(b)}_{kj} = \gamma^{(b-1)}_{kj}$ otherwise.

\subsubsection{Pseudo-values.} \label{subsub:pseudovalues}
Following \cite{andersen2003},
\citet{klein2005} propose a method based on the jackknife (leave-one-out) CIF estimator 
and GLMs. 
This is similar to the \citet{Fine2001} method, which extends the model in \eqref{eq:CIF_reg} to use an arbitrary link function. Given a time point grid, $\tau_1, \ldots, \tau_M$, \cite{andersen2003} define {\it pseudo-values} for the CIF of the $i$-th individual at time point $\tau_m$ for the $k$-th event type. These are given by 
\begin{equation}
    \theta_{imk} = n \CIF_{k}(\tau_m) - (n-1) \CIF^{ -i}_{k}(\tau_m),
\end{equation}
where $\CIF_{k}(\tau_m)$ is the Aalen-Johansen \citep{AJ1978} CIF estimator evaluated 
using all the data 
and $\CIF^{ -i}_{k}(\tau_m)$ is the corresponding estimate after removing the $i$-th observation. Then, based on these pseudo-values, a GLM is used to estimate covariate effects on the CIF:
\begin{equation}
    g(\theta_{imk}) = \alpha_{mk} + \bx_i^\top \bm{\gamma}_k, \label{eq:pseudo}
\end{equation}
where $\alpha_{mk}$ and $ \bm{\gamma}_k=(\gamma_{k1},\ldots, \gamma_{kp})^\top$ are regression coefficients estimated via generalised estimating equations \citep{GEE} and $g(\cdot)$ is a link function (for the 
complementary log-log link, 
\eqref{eq:CIF_reg} is recovered). Users select $\tau_1, \ldots \tau_M$ a priori and the authors suggest five to ten equally spaced points for this purpose.
Furthermore, time-varying covariate effects can be added via: 
\begin{equation} \label{eq:pseudo_timevarying}
    g(\theta_{imk}) = \alpha_{mk} + \bm{v}_i^\top \bm{\eta}_k(t_i) +\bm{u}_i^\top \bm{\gamma}_k,
\end{equation} 
where observed covariates, $\bx_i$, are split into those with time varying effects ($\bm{v}_i$) and those with constant effects ($\bm{u}_i$), whose corresponding regression coefficients are $\bm{\eta}_k(t_i)$ and $\bm{\gamma}_k$, respectively.


\subsubsection{Direct binomial.}
\cite{Scheikeetal2008} propose a semi-parametric strategy, extending \eqref{eq:CIF_reg} to a more general class that enables both, time-varying and constant covariate effects. 
This includes a goodness-of-fit test to check if time-varying effects are required. 
The CIF is defined as
\begin{equation}\CIF_k(t_i \mid \bx_i)= g^{-1}(\bm{\eta}_k(t_i), \bm{\gamma}_k, \bx_i), \end{equation} where $g(\cdot)$ is a known link function, $\bm{\eta}_k(t_i)$ are time varying parameters, and $\bm{\gamma}_k$ captures constant covariate effects. Both, $\bm{\gamma}_k$ and $\bm{\eta}_k(t_i)$ are estimated through score equations.
More precisely, \cite{Scheikeetal2008} studied an additive and multiplicative specification, defined respectively as
\begin{align}
     g(\CIF_k(t_i \mid \bx_i)) & = {\bm{v}_i}^\top \bm{\eta}_k(t_i) + f( \bm{\gamma}_k, \bm{u}_i, t_i), \hspace{1cm} \text{and} \\
    g(\CIF_k(t_i \mid \bx_i)) & = \left[ {\bm{v}_i}^\top \bm{\eta}_k(t_i) \right] f( \bm{\gamma}_k, \bm{u}_i, t_i),
\end{align}
where $f(\cdot)$ is a known function and $\bx_i$ is split as in \eqref{eq:pseudo_timevarying}. 


\subsubsection{Parametric constrained CIF.} \cite{shi2013} extended \eqref{eq:CIF_reg} as 
$$g_k(\CIF_k(t_i \mid \bx_i)) = g_k(\CIF_{k0}(t_i)) +  \bx^\top_{i} \bm{\gamma}_k, \quad k=1,2; $$
where the link functions $g_k(\cdot)$ are the
generalised odds rate model by \cite{jeong2006}: 
\begin{equation}
    g_k(u) = \log \left[ \{ (1-u)^{-\alpha_k} -1\}/\alpha_k\right], \quad \text{with} \quad 0<\alpha_k<\infty. \label{eq:OR}
\end{equation}
To ensure that $\CIF_1(t \mid \bx_i) + \CIF_2(t \mid \bx_i) = 1$ as $t \rightarrow \infty$, \cite{shi2013} treat both events differently. For the primary event ($k=1$), $\CIF_1(t)$ is set using a modified three-parameter logistic function \citep{Cheng2009} for the baseline hazard:
\begin{equation} \label{eq:constrainedCIF_baseline}
 \CIF_{10}(t_i) = \frac{p_1 [ \exp \{ b_1(t_i-c_1)- \exp(-b_1 c_1)\}]}{1+\exp\{ b_1(t_i-c_1)\}},
\end{equation}
where $p_k$ is the log-term probability of the $k$-th event  ($\CIF_k(t) \rightarrow p_k$ as $t \rightarrow \infty$), $b_k >0$ dictates how fast $\CIF_k(t)$ approaches $p_k$, and $c_k \in \mathbb{R}$. 
Instead, for the competing event ($k=2$), they do not specify direct covariate effects and the CIF is given by: 
\begin{equation}
     \CIF_{2}(t_i \mid \bx_i) = \frac{p_2(\bx_i)[ \exp \{ b_2(t_i-c_2)- \exp(-b_2 c_2)\}]}{1+\exp\{ b_2(t_i-c_2)\}},
\end{equation}
with $p_2(\bx_i) = (1-p_1)^{\exp(\bx^\top_{i} \bm{\gamma}_1)}$ and, where $b_2$ and $c_2$ as as in \eqref{eq:constrainedCIF_baseline}.
Inference is performed via maximum likelihood, and can can be extended to allow for right,
interval and left censoring. 

\subsubsection{Dependent Dirichlet processes (DDP).} \citet{shi2021dependent} introduced a Bayesian non-parametric approach for $K=2$ based on the specification by \cite{Fan2008} and infinite mixtures of Weibull distributions \citep{KOTTAS2006}. First, whilst $\CIF_1(t \mid \bx_i)$ is defined as in \eqref{eq:CIF_reg}, the second CIF is modified to ensure that $\CIF_1(t \mid \bx_i) + \CIF_2(t \mid \bx_i) = 1$ as as $t \rightarrow \infty$. Secondly, baseline CIFs are parametrized in terms of \emph{normalized} distributions as \begin{equation}
\CIF_{10}(t) = c \times D_{01}(t) \text{\hspace{0.5cm} and \hspace{0.5cm}} \CIF_{20}(t) = (1-c) \times D_{02}(t),   
\end{equation} where the normalising constant is given by $c = \lim_{t \rightarrow \infty} \CIF_{10}(t)$. This leads to \begin{equation}
\CIF_2 (t_i \mid \bx_i) = (1-c)^{\exp(\bx_i^\top \bm{\gamma}_1)} \left(1-(1-D_{02}(t_i))^{\exp(\bx_i^\top \bm{\gamma}_2)} \right).     
\end{equation} Finally, assuming that $D_{01}(t)$ and $D_{02}(t)$ correspond to Weilbull distributions, the DDP model defines the $i$-th subject likelihood contribution as a Dirichlet Process mixture model \citep{escobar1995} which permits clustering of observations.  More details about the mixture model including prior and hyperparameter choices are discussed in \citet[Section 6]{shi2021dependent}.
The DDP approach scales linearly with the sample size as well as with the number of features. Moreover, it permits inference with interval censored data and time-dependent covariates.

\subsubsection{Survival Multitask Boosting (SMTBoost).} 
SMTBoost \citep{Bellot2018} is a non-parametric method that combines boosting and multi-task learning \citep{Caruana93multitasklearning} to jointly estimate the CIF associated to all event types. The aim is to minimize the difference between the observed and predicted survival status via the following loss function: \begin{equation}
\tfrac{1}{K}\sum_{k=1}^{K} \E\left[  \tfrac{1}{\tau} \int_0^{\tau}( \mathbbm{1}\{ T_i \leq t, Z_i=k\} - \CIF_k(t\mid \bx_i))^2 dt \right]    
\end{equation} SMTBoost uses binary partitioned trees as weak learners and 
a modification of Gray’s test \citep{gray1988} for the splitting rule. 
An iterative boosting procedure is implemented by re-weighting samples based on their prediction error.
At each terminal 
node $m$, the Aalen-Johansen estimator \citep{AJ1978} for the $k$-th CIF, $\CIF_{k,m}^{\text{AJ}}(t_i)$, is calculated. Let $\mathcal{C}_m$ denote the index set of observations in 
node $m$,  
the 
CIF is then computed as
\begin{equation}
   \CIF_k(t_i \mid \bx_i) = \sum_m \mathbbm{1} \{  i \in \mathcal{C}_m\} \CIF_{k,m}^{\text{AJ}}(t_i). 
\end{equation}


SMTBoost permits covariate selection via a variable importance measure computed per event type. The authors showed good performance for event types with low incidence, 
in datasets with a large number of observations as well as high-dimensional covariate spaces. 

\subsubsection{Derivative-based neural network modelling (DeSurv).} 
\cite{danks2022derivative} proposed a flexible, non-parametric approach that can be seen as a continuous time version of the work by \citet{lee2018deephit} (DeepHit). The method can be applied in simple survival settings (only one event type) and in the presence of CR. DeSurv for CR, decomposes the CIF in~\eqref{eq:CIF} as \begin{eqnarray}
    \CIF_k(t_i \mid \bx_i) & = & \Pr(T \leq t_i \mid  Z_i = k , \bx_i ) \Pr(Z_i = k \mid \bx_i ) \nonumber \\
    & \equiv & \tilde{F}_k(t_i \mid \bx_i) \pi_{ik}(\bx_i), \hspace{0.2cm} \text{with} \hspace{0.2cm} \sum_{k=1}^K \pi_{ik}(\bx_i) = 1.
\end{eqnarray} Note that this factorisation is similar to the one used by the mixture models in Section \ref{S: mixtures}. DeSurv models $\pi_{ik}(\bx_i)$ using a neural network with a softmax activation function and $\tilde{F}_k(t_i \mid \bx_i)$ using the DeCDF approach \citep[][Section 2.2]{danks2022derivative} which assumes it is as a monotonic transformation of a function whose derivative is parametrised as a neural network. 
In this setting, training is performed using the Adam optimisation algorithm\citep{Kingma2015Adam}.

\subsection{Approaches based on a latent survival times specification}

\subsubsection{Deep multi-task Gaussian Processes.} \label{subsec:DMTGP}
\cite{Alaa2017deep} proposed a Bayesian non-parametric method using deep Gaussian Processes (GPs) \citep{damianou13a}. 
The approach permits to estimate 
patient-specific survival times as well as the CIFs.
Unlike other neural networks-based approaches for CRs, deep multi-task GPs have a fully probabilistic flavour. This is based on the following hierarchical construction: \begin{align}
      \bm{T}_i | \bm{\zeta}_i &\sim \Norm \left (f_T(\bm{\zeta}_i), \omega^2_T\bm{I}_K \right), \label{eq:DGP_eq1}\\
      \bm{\zeta}_i &\sim \Norm \left( f_{\zeta}(\bx_i), \omega^2_{\zeta}\bm{I}_q \right). 
 \end{align} 
where $\bm{T}_i = (T_{i1}, \ldots, T_{iK})^\top$ is a vector of latent survival times for subject $i$, $\bm{I}_q$ is a $q$-dimensional identity matrix and $\bm{\zeta}_i$ a $q$-dimensional latent variable ($q=3$ was used in the original publication). Moreover, 
$f_T(\cdot): \mathbb{R}^q \rightarrow \mathbb{R}^{K}$ and $f_{\zeta}(\cdot): \mathbb{R}^p \rightarrow \mathbb{R}^{q}$  are independent zero centred vector-valued GPs whose covariance functions $C_{\theta_\zeta}(\cdot, \cdot)$ and $C_{\theta_{T}}(\cdot, \cdot)$ depend on parameters $\theta_{\zeta}$ and $\theta_{T}$, respectively. These are defined using the intrinsic coregionalization model \citep[][Section 4.2]{alvarez2012}. 
Inference is performed 
with a variational framework \citep{blei2017} 
that can be combined with the inducing points approach of \cite{titsias2009} 
to derive a tractable algorithm. 
Note that \eqref{eq:DGP_eq1} assumes independence among the latent survival times (conditional on $\bm{\zeta}_i$).



\subsubsection{Bayesian Lomax delegate racing (LDR).} 
LDR \citep{zhang2018nonparametric} can be seen as a generalisation of exponential racing \citep{caron2012} 
or as a Gamma process \citep{Wolpert1998}. It assumes that every event type is determined by a potentially infinite number of sub-risks (e.g.~different etiologies of a disease). To facilitate implementation, \cite{zhang2018nonparametric} truncated the Gamma process to use $L$ sub-risks ($L=10$ was used in their experiments). 
LDR relaxes the PH assumption, allows non-linear covariate effects and for missing event times or types. 
Let $T_{ik} = \min\{T_{ik1}, \ldots, T_{ikL}\}$ and ${\bx_i^\prime} = (1, x_{i1}, \ldots, x_{ip})^{\top}$, the LDR model is based on the following hierarchical formulation: 
\begin{align}
  T_{ikl} | \lambda_{ikl} &\sim \Exp(\lambda_{ikl} \exp( {\bx_i^\prime}^\top \bm{\nu}_{kl})) \nonumber \\
  \lambda_{ikl} | \alpha_{kl}, \bm{\nu}_{kl}  &\sim \Gammad \left(\alpha_{kl}, 1 \right), 
\end{align}
 where $\bm{\nu}_{kl} = (\nu_{kl0},\ldots, \nu_{klp})^\top$ are regression parameters for the $k$-th event. Prior distributions for $\alpha_{kl}$ and $\bm{\nu}_{kl}$ 
 are discussed in \citet[][Appendix B]{zhang2018nonparametric}.  
After marginalisation of $\lambda_{ikl}$, inference uses a Gibbs sampler \citep{geman1984} for moderate $n$, 
and 
maximum a posteriori 
through stochastic gradient descent \citep{kiefer1952stochastic} for larger datasets.  LDR assumes conditional independence across cause-specific latent survival times. 

\subsection{Others}

In this section we introduce methods that do not fall in any of the previous categories or that can accommodate more than one specification.
For instance, 
\cite{Ishwaran2014} enable, both, a CS hazard and a CIF formulation for covariate effects.

\subsubsection{Mixture models.} \label{S: mixtures} 
These models decompose the joint distribution of the event time and event type into 
marginal probabilities 
$\pi_{ik}(\bx_i) = \Pr(Z_i = k \mid \bx_i )$ of each event type (mixing proportions) and the conditional survival distribution $S_k(t_i \mid \bx_i)= \Pr(T > t_i\mid Z_i=k, \bx_i)$. This assumes that each individual will experience an specific event type, which is chosen randomly \citep{larson1985}. 
Mixture models 
typically need large sample sizes to avoid identifiability issues \citep{haller2013}. 
In general, the CR model is set as a $K$-component mixture 
\begin{equation}S(t_i \mid \bx_i) = \sum_{k=1}^{K} \pi_{ik}(\bx_i)S_k(t_i \mid \bx_i), \hspace{0.5cm} \text{with} \hspace{0.5cm} \sum_{k=1}^{K} \pi_{ik}(\bx_i) = 1.\label{eqref:mixture_surv}\end{equation}

In this context, several model specifications have been proposed. In particular, \cite{larson1985} 
assumed the number of events across types to be multinomial with probabilities: 
\begin{equation} \label{eq: mixture_weights}
    \pi_{ik}(\bx_i) = \frac{\exp \left( a_k + \bx_i^\top \bm{b}_k \right)}{1+ \sum_{l=1}^{K-1} \exp \left(a_l + \bx_i^\top \bm{b}_l \right) }, \hspace{1cm} k=1, \ldots, K,
\end{equation}
where $a_k$ and $\bm{b}_{k}$ are regression coefficients.
In addition, they assume\begin{equation}
   S_k(t_i \mid \bx_i) = \exp \left[ - \int_0^{t_i} h_{k0}(u) \exp( \bx^\top_{i} \bm{\theta}_{k}) \,du \right],
\end{equation}
where the baseline hazards, $h_{k0}(t_i)$, are
piecewise constant functions 
within $L$ disjoint intervals. 
The formulation above implicitly introduces covariate effects via the CIF through the following identity $\CIF_k(t_i \mid \bx_i) = \pi_{ik}(\bx_i) \left(1 - S_k(t_i \mid \bx_i) \right)$. 

Maximum likelihood estimates 
can be obtained using expectation-maximisation \citep[EM,][]{Dempster1977}. One challenge is to select 
$L$: a large $L$ 
may lead to an overparametrised model; a small $L$ may cause poor fitting \citep{kuk1992}. 
To overcome this, \cite{kuk1992} propose a semi-parametric model with arbitrary baseline hazards 
and suggests to infer $a_k,$ $\bm{b}_k$ and $\bm{\theta}_k$ 
using a Monte Carlo approximation of the marginal likelihood. 
These estimates are subsequently used within EM to infer 
$h_{k0}(t_i)$.
Similarly, \cite{ng2003based} propose a semi-parametric approach 
that uses expectation-conditional maximisation \citep{ECM1993}. 
 In this case, multiple initialisations may be required to ensure convergence of the algorithm. 
Furthermore, \cite{Chang2007Nonparametric} propose a different algorithm for maximum likelihood estimation along with asymptotic properties of the estimators.

Finally, \cite{TMixtureM2018} introduced a tree-based mixture model based on a Bayesian semi-parametric method that uses generalised gamma distributions  \citep{cox2007} to model the conditional
survival distributions in \eqref{eqref:mixture_surv}, such that $S_k(t_i \mid \bx_i) := \GG(t_i \mid \theta_{i k}, \sigma_k, \lambda_k)$, where $\theta_{ik}$ is
a subject and CS parameter, and $\sigma_k >0, \lambda_k$ are only cause specific. 
The vectors $\bm{\theta}_i := (\theta_{i 1}, \ldots \theta_{i K})^\top$ and $\bm{\pi}_i := (\pi_{i 1}, \ldots \pi_{i K})^\top$ are defined as
\begin{equation}
    \begin{aligned}
   \bm{\theta}_i \mid \bx_i = g_\theta(\bx_i) + \epsilon_{\beta i},& \quad  \epsilon_{\theta i} \sim \Norm(0, \omega^2_\theta) \\
 \bm{\pi}_i \mid \bx_i = l(g_\pi(\bx_i) + \epsilon_{\pi i}),&  \quad  \epsilon_{\pi i},\sim \Norm(0, \omega^2_\pi) ,
\end{aligned}
\end{equation}
with $g_\theta$ and $g_\pi$ functions $\mathbb{R}^p \rightarrow \mathbb{R}^K$ modelled with Multivariate Random Forests \citep{segal2011}, $l(x_i) = x_i/ \sum_i x_i $, and $(\omega^2_\theta, \omega^2_\pi)$ fixed hyperparameters.  Gamma and Gaussian priors are assigned to $\sigma_k$ and $\lambda_k$, respectively. 
Inference is done with 
an adaptive Metropolis-within-Gibbs scheme \citep{hastings1970}. 
 This model can handle non-linear covariate effects, provides information about variable importance for each event type, and permits to infer individualised survival estimates. However, the current implementation does not scale well for large $n$.


\subsubsection{Vertical modelling.} 
\cite{Nicolaie2010} propose to decompose the joint distribution of the event time and type 
to first estimate the overall probability of event occurrence and then the probability of a specific event type given that the event occurred at a given time.
This decomposition is unlike the one used by other mixtures models (Section~\ref{S: mixtures}), 
which are formulated in the opposite manner. 
The vertical modelling approach requires to fit two models, one for the overall hazard function 
and one for the relative CS hazard defined as $r_k(t_i) =  {h_k(t_i)}/{h(t_i)}$.

The overall hazard function $h(t)$ can be estimated using a PH approach or with a Nelson-Aalen estimator for a single categorical covariate. In this case, all event types are considered as events, regardless of their cause. Instead, the CS relative hazard can be fitted with a multinomial logistic regression combined with spline basis functions to smooth the function over time.  \cite{Nicolaie2010} discussed two specifications for $r_k(t_i \mid \bx_i)$, one that incorporates interaction effects and the other with an additive structure. The latter is given by
\begin{equation}
r_k(t_i \mid \bx_i) = \frac{\exp \left(  \bm{b}(t_i)^\top \bm{\eta}_k + \bx^\top_{i} \bm{\theta}_{k} \right)}{\sum_{l=1}^K \exp( \bm{b}(t_i)^\top \bm{\eta}_l+ \bx^\top_{i} \bm{\theta}_{l})}, \label{eq: relative_h}
\end{equation}
where $\bm{\theta}_k$ 
denotes a $p$-dimensional vector of 
covariate effects, $\bm{b}(t_i)$ introduces smooth dependency on $t$ via $q$ spline basis functions 
and $\bm{\eta}_k$ 
represents the regression coefficients associated to them. For identifiability, all entries of 
$\bm{\theta}_1$ and $\bm{\eta}_1$ are set equal to one. 
The 
model in \eqref{eq: relative_h} can be extended to allow for interactions between covariates and splines.
While interpretability of the regression coefficients in the relative CS hazard can be challenging, a graphical representation of the estimated relative hazards over time can provide relevant insights. For instance, one can infer the contribution of the different event types to the overall rate of failure along time. 
Note that this approach is implicitly modelling covariate effects via a CS hazard, $h_k(t)$; however, the method does not follow a PH assumption.
\cite{Nicolaie2015} have extended this approach to deal with missing event types. 



\subsubsection{Random survival forests (RSF).}
\cite{Ishwaran2014} introduced a non-parametric approach using 
an ensemble of random forests \citep{breiman2001}. RSF can handle high-dimensional and large data problems, several competing events, and permits non-linear/interaction covariate effects. 
%
RSF uses a pre-fixed number of boostrap samples ($B$) to grow $B$ trees using a random selection of covariates at each node. During the tree construction, each node is divided based on a splitting criterion and the tree is grown to meet a specific stopping rule. 
More specifically, two event-specific splitting rules are proposed: one based on the log-rank test \citep{mantel1966evaluation} and one related to a modification of the Gray’s test \citep{gray1988}. The former, is better suited to select covariates that influence a CS hazard; the latter, to test covariate effects on the CIF. In addition, a splitting rule that combines the two 
rules is proposed for cases where the objective is to select covariates that affect any cause or when the goal is to predict the CIF of all causes. The splitting rules are detailed in \citet[][Section 3.3]{Ishwaran2014}.

For each tree, RSF computes the $k$-th CIF using the Aalen-Johansen estimator 
and a measure of the expected number of time lost due to the $k$-th cause before time $\tau$: 
\begin{equation}
M^{(b)}_{k}(\tau \mid \bx_i) = \int_0^\tau \CIF_{k}^{(b)\text{AJ}}(t_i \mid \bx_i)\,dt,    
\end{equation}
where $\tau$ is such that the probability of being uncensored is bounded away from zero.
Overall estimates are reported 
by averaging over the trees. Finally, RSF can perform variable selection via an event-specific variable importance \citep[Section 5]{Ishwaran2014} and a non-event-specific minimal depth metric \citep{Ishwaranvarselection}. 

Note that \cite{Janitza2015prediction} proposed another approach based on random forests using a discrete scale for the survival times.

\subsubsection{Deep Survival machines (DSM).}
\cite{Nagpal2021} propose a model that is similar to DeepHit \citep{lee2018deephit}, but uses a continuous-time approach that is better suited for long survival horizons. For a single event type, the distribution of the survival times is defined as a finite mixture of parametric distributions, either Weibull or log-normal. The parameters of each mixture component are set as a non-linear function of the covariates via a neural network which, in turn, learns a distributional representation for the input covariates. 
Both, the network and survival parameteres, 
are jointly learned during the training procedure by optimising a loss function which down-weights the contribution of censored observations to reduce potential biases towards long-tails in the survival distribution. 
In their experiments, the authors use cross-validation to tune hyperparameter choices (e.g. the number of mixture components).  
When extending their approach to address CR, \cite{Nagpal2021} use a single neural network to learn a common representation for input covariates that is shared across event types. During training, competing events are treated as censored observations.

\section{Competing risks survival models for discrete time-to-event data} \label{S:DiscreteCR}

So far, the models included in this review focus on continuous survival times. However, time-to-event outcomes are often recorded in a discrete scale (e.g. weeks, months). Recently, \cite{discretereview} provided an overview for approaches developed in this context. The predominant method is a CR extension for the \emph{proportional odds} model \citep{Cox1972}. This introduces covariate effects through a discrete-time version of the cause-specific hazard function: \begin{equation} \label{eq:discreteHazard}
    h_k^{\text{D}}(t) = \frac{\text{Pr}(T = t, Z = k)}{\text{Pr}(T \geq t)}.
\end{equation} The CR proportional odds model \citep{tutz1995competing} is then defined as: \begin{equation} 
\log \left( \frac{h_k^{\text{{D}}}(t_i \mid \bx_i)}{h_0^{\text{{D}}}(t_i \mid \bx_i)} \right) = \lambda_{kt_{i}} + \bx_i^\top\bm{\omega}_k, \hspace{0.5cm} k = 1, \ldots, K, 
    \label{eq:prop_odds}
\end{equation} where $h_0^{\text{{D}}}(t_i \mid \bx_i) = 1 - \sum_{k=1}^K h_k^{\text{{D}}}(t_i \mid \bx_i)$, $\lambda_{kt_{i}}$ are baseline log-odds ($k$-th event versus no event) and $\bm{\omega}_k = (\omega_{k1}, \ldots, \omega_{kp})^\top$ regression coefficients. This model can be estimated in most statistical software using multinomial logistic regression. For this purpose, the data is transformed into a person-period format \citep{scott2005}, using binary indicators $Y_{itk} = \mathbbm{1}{ \{T_i = t, Z_i= k \}}$ to capture whether an event of type $k$ is observed at time $t$ for individual $i$. In this context, the CIF for each event type can be then estimated as \begin{equation}
   \CIF_k(t_i\mid \bx_i) = \sum_{t=0}^{t_i} q(t, k \mid \bx_i), \label{eq: discrete CIF}
\end{equation} where $q(t, k \mid \bx_i) = \text{Pr}(T_i = t, Z_i = k \mid \bx_i)$. 

The person-period representation of discrete time CR datasets has enabled several extensions for the model in \eqref{eq:prop_odds} based on statistical and machine learning approaches developed for binary or multinomial outcomes \citep[see][for an overview]{discretereview}. Examples include SSPN \citep{Siames2018} and DeepHit \citep{lee2018deephit}, both using neural networks. Another recent approach, by \cite{Bart2020}, is based on Bayesian additive regression trees \citep[BART,][]{reviewbart}. BART permits non-linear/interaction effects, non-proportional hazards, missing data and uses a sparse prior for high-dimensional covariate spaces. For completeness, as the method by \cite{Bart2020} is implemented within the popular BART R package \citep{rBART}, we decided to include it in this review.

\citeauthor{Bart2020} assume that the binary indicators $Y_{itk}$ follow a multinomial distribution with event probabilities $\pi_{itk} = \Pr(T_i= t, Z_i= k \mid T_i \geq t, \bx_i)$, which can be seen as a discrete hazard (if the survival times are not discrete, a discretised scale is adopted with each observed/censored time treated as a distinct time-point). As multinomial implementations of BART are not widely available, the authors propose two formulations using BART probit models, focusing on $K=2$. In the fist formulation, one 
model is used for the time until \emph{any} event occurs and a second 
model for the conditional probability of the event being of type $k=1$ given that an event occurred. In contrast, the second formulation employs one model
for the conditional probability of experiencing event type $k=1$ at time $t$ given that the subject is still at risk. A second model is then used for the conditional probability of a type $k=2$ event at time $t$ given that the subject is still at risk and that it has not experience a type $k=1$ event. Prior distributions for the required parameters in the models are discussed in detail in \citet[Section 2]{Bart2020}.
 \begingroup
 \renewcommand{\arraystretch}{.8}
\begin{table}[htbp]
\centering
\begin{threeparttable}
\caption{Summary of the available methods for survival regression with CR.}
  \centering
  \small
   {\tabcolsep=4.25pt
   \begin{tabular}{lcllll}
    \toprule
    \multicolumn{1}{c}{\multirow{2}[1]{*}{Model}}  &    
    \multicolumn{1}{c}{\multirow{2}[1]{*}{Type}}   & 
    \multicolumn{1}{c}{High} &
    \multicolumn{1}{c}{Non-linear} &
    \multicolumn{1}{c}{Missing} & 
    \multicolumn{1}{c}{Covariate}  \\
          &           &         
    \multicolumn{1}{c}{dimensions $(p)$} & 
    \multicolumn{1}{c}{effects} &      
    \multicolumn{1}{c}{data} & 
    \multicolumn{1}{c}{interpretability}   \\
    \midrule
    \multicolumn{6}{l}{\bf Approaches based on a proportional cause-specific hazard specification} \\
    \midrule
      \multicolumn{1}{c}{\multirow{1}[1]{*}{Cox proportional}} &
      \multicolumn{1}{c}{\multirow{2}[1]{*}{Semi-parametric}}& 
      \multicolumn{1}{c}{\multirow{2}[1]{*}{\ding{55}}} & \multicolumn{1}{c}{\multirow{2}[1]{*}{\ding{55}}} &
     \multicolumn{1}{c}{\multirow{2}[1]{*}{\ding{55}}} & \multicolumn{1}{c}{\multirow{2}[1]{*}{\ding{51}}}  \\
      \multicolumn{1}{c}{\multirow{1}[1]{*}{cause-specific hazard}} &        &       &       &     
     & \\
        \multicolumn{1}{c}{\multirow{2}[0]{*}{}}  &         
    &  &       &       
    &       \\
      \multicolumn{1}{c}{\multirow{1}[1]{*}{Penalised Cox}} &
      \multicolumn{1}{c}{\multirow{2}[1]{*}{Semi-parametric}}& 
      \multicolumn{1}{c}{\multirow{2}[1]{*}{\ding{51}}} & \multicolumn{1}{c}{\multirow{2}[1]{*}{\ding{55}}} &
     \multicolumn{1}{c}{\multirow{2}[1]{*}{\ding{55}}} & \multicolumn{1}{c}{\multirow{2}[1]{*}{\ding{51}}}  \\
      \multicolumn{1}{c}{\multirow{1}[1]{*}{proportional hazard}} &        &       &       &       
      &    \\
        \multicolumn{1}{c}{\multirow{2}[0]{*}{}}  &         
    &  &       &       
    &       \\
     \multicolumn{1}{c}{\multirow{1}[1]{*}{Cox model-based}} &
      \multicolumn{1}{c}{\multirow{2}[0]{*}{Semi-parametric}} & 
      \multicolumn{1}{c}{\multirow{2}[0]{*}{\ding{51}}}&  \multicolumn{1}{c}{\multirow{2}[0]{*}{\ding{55}}}&   
     \multicolumn{1}{c}{\multirow{2}[0]{*}{\ding{55}}}&   \multicolumn{1}{c}{\multirow{2}[0]{*}{\ding{51}}} \\
     \multicolumn{1}{c}{\multirow{1}[1]{*}{boosting}}  &  &   & 
        &      
        &   \\
          \multicolumn{1}{c}{\multirow{2}[0]{*}{}}  &         
    &  &       &      
    &       \\
      \multicolumn{1}{c}{\multirow{1}[1]{*}{Cox likelihood-based}}&
       \multicolumn{1}{c}{\multirow{2}[0]{*}{Semi-parametric}}&
       \multicolumn{1}{c}{\multirow{2}[0]{*}{\ding{51}}} &  \multicolumn{1}{c}{\multirow{2}[0]{*}{\ding{55}}}& 
       \multicolumn{1}{c}{\multirow{2}[0]{*}{\ding{55}}}&    \multicolumn{1}{c}{\multirow{2}[0]{*}{\ding{51}}}     \\
           \multicolumn{1}{c}{\multirow{1}[1]{*}{boosting}}  &  &  
               &       &     
               &   \\
    \multicolumn{1}{c}{\multirow{2}[0]{*}{}}  & & & & & \\
     \multicolumn{1}{c}{\multirow{2}[0]{*}{Lunn-McNeil}} &  
     \multicolumn{1}{c}{\multirow{2}[0]{*}{Semi-parametric}}&  
     \multicolumn{1}{c}{\multirow{2}[0]{*}{ \ding{55} }} & 
     \multicolumn{1}{c}{\multirow{2}[0]{*}{ \ding{55} }} & 
     \multicolumn{1}{c}{\multirow{2}[0]{*}{ \ding{55} }} & 
     \multicolumn{1}{c}{\multirow{2}[0]{*}{ \ding{51} }}  \\
     \multicolumn{1}{c}{} &        
      &  &       &      
      &       \\
       \midrule 
    \multicolumn{6}{l}{\bf Approaches based on the CIF} \\
    \midrule     
\multicolumn{1}{c}{\multirow{2}[0]{*}{Fine-Gray}}& 
      \multicolumn{1}{c}{\multirow{2}[0]{*}{Semi-parametric}}&
     \multicolumn{1}{c}{\multirow{2}[0]{*}{\ding{55}}} & \multicolumn{1}{c}{\multirow{2}[0]{*}{\ding{55}}} & 
     \multicolumn{1}{c}{\multirow{2}[0]{*}{\ding{55}}} & \multicolumn{1}{c}{\multirow{2}[0]{*}{\ding{51}}}   \\
     \multicolumn{1}{c}{\multirow{2}[0]{*}{}}  &         
    &  &       &      
    &       \\
          \multicolumn{1}{c}{\multirow{1}[1]{*}{Penalised proportional}} &
      \multicolumn{1}{c}{\multirow{2}[1]{*}{Semi-parametric}}& 
      \multicolumn{1}{c}{\multirow{2}[1]{*}{\ding{51}}} & \multicolumn{1}{c}{\multirow{2}[1]{*}{\ding{55}}} &
      \multicolumn{1}{c}{\multirow{2}[1]{*}{\ding{55}}} & \multicolumn{1}{c}{\multirow{2}[1]{*}{\ding{51}}}  \\
      \multicolumn{1}{c}{\multirow{1}[1]{*}{sub-distribution hazard}} &        &       &       &       
      &    \\
      \multicolumn{1}{c}{\multirow{2}[0]{*}{}}  & & & & & \\
      \multicolumn{1}{c}{\multirow{1}[1]{*}{Sub-distribution}}&
       \multicolumn{1}{c}{\multirow{2}[0]{*}{Semi-parametric}}&
       \multicolumn{1}{c}{\multirow{2}[0]{*}{\ding{51}}} &  \multicolumn{1}{c}{\multirow{2}[0]{*}{\ding{55}}}& 
       \multicolumn{1}{c}{\multirow{2}[0]{*}{\ding{55}}}&    \multicolumn{1}{c}{\multirow{2}[0]{*}{\ding{51}}}     \\
           \multicolumn{1}{c}{\multirow{1}[1]{*}{hazard boosting}}  &  &  
               &       &     
               &   \\
      \multicolumn{1}{c}{\multirow{2}[0]{*}{}}  & & & & & \\  
        \multicolumn{1}{c}{\multirow{2}[0]{*}{Pseudo-values}} & 
    \multicolumn{1}{c}{\multirow{2}[0]{*}{Semi-parametric}} &  
    \multicolumn{1}{c}{\multirow{2}[0]{*}{\ding{55}}} &   
    \multicolumn{1}{c}{\multirow{2}[0]{*}{\ding{55}}} & 
    \multicolumn{1}{c}{\multirow{2}[0]{*}{\ding{55}}} &  \multicolumn{1}{c}{\multirow{2}[0]{*}{\ding{51}}}  \\   \multicolumn{1}{c}{\multirow{2}[0]{*}{}}  &         
    &  &       &     
    &       \\
     \multicolumn{1}{c}{\multirow{2}[0]{*}{Direct binomial}} &  
    \multicolumn{1}{c}{\multirow{2}[0]{*}{Semi-parametric}}& 
    \multicolumn{1}{c}{\multirow{2}[0]{*}{\ding{55}}} & \multicolumn{1}{c}{\multirow{2}[0]{*}{\ding{51}}} & 
   \multicolumn{1}{c}{\multirow{2}[0]{*}{\ding{55}}} & \multicolumn{1}{c}{\multirow{2}[0]{*}{\ding{51}}} \\
     \multicolumn{1}{c}{\multirow{2}[0]{*}{}}  &         
    &  &       &    
    &       \\

     \multicolumn{1}{c}{\multirow{1}[1]{*}{Parametric}} &
      \multicolumn{1}{c}{\multirow{2}[1]{*}{Parametric}}& 
      \multicolumn{1}{c}{\multirow{2}[1]{*}{\ding{55}}} & \multicolumn{1}{c}{\multirow{2}[1]{*}{\ding{51}}} &
      \multicolumn{1}{c}{\multirow{2}[1]{*}{\ding{55}}} & \multicolumn{1}{c}{\multirow{2}[1]{*}{\ding{51}}}  \\
      \multicolumn{1}{c}{\multirow{1}[1]{*}{constrained CIF}} &        &       &       &   
      &    \\
        \multicolumn{1}{c}{\multirow{2}[0]{*}{}}  &         
    &  &       &   
    &       \\
    \multicolumn{1}{c}{\multirow{2}[0]{*}{Dependent DP}} &  
    \multicolumn{1}{c}{\multirow{2}[0]{*}{Non-parametric}} &  
    \multicolumn{1}{c}{\multirow{2}[0]{*}{\ding{55}}} &   
    \multicolumn{1}{c}{\multirow{2}[0]{*}{\ding{51}}} & 
    \multicolumn{1}{c}{\multirow{2}[0]{*}{\ding{55}}} &  \multicolumn{1}{c}{\multirow{2}[0]{*}{\ding{51}}}  \\   \multicolumn{1}{c}{\multirow{2}[0]{*}{}}  &         
    &  &       &   
    &       \\
 \multicolumn{1}{c}{\multirow{2}[0]{*}{SMTBoost}}  & 
     \multicolumn{1}{c}{\multirow{2}[0]{*}{Non-parametric}} 
    & 
    \multicolumn{1}{c}{\multirow{2}[0]{*}{\ding{51}}} &   \multicolumn{1}{c}{\multirow{2}[0]{*}{\ding{51}}} &  
   \multicolumn{1}{c}{\multirow{2}[0]{*}{\ding{55}}} & 
    \multicolumn{1}{c}{\multirow{2}[0]{*}{\ding{55}}} \\
      \multicolumn{1}{c}{\multirow{2}[0]{*}{}}  &         
    &  &       &     
    &       \\
  \multicolumn{1}{c}{\multirow{2}[0]{*}{DeSurv}}  & 
     \multicolumn{1}{c}{\multirow{2}[0]{*}{Non-parametric}} 
    & 
    \multicolumn{1}{c}{\multirow{2}[0]{*}{\ding{55}}} &   \multicolumn{1}{c}{\multirow{2}[0]{*}{\ding{51}}} &  
   \multicolumn{1}{c}{\multirow{2}[0]{*}{\ding{55}}} & 
    \multicolumn{1}{c}{\multirow{2}[0]{*}{\ding{55}}} \\
      \multicolumn{1}{c}{\multirow{2}[0]{*}{}}  &         
    &  &       &     
    &       \\   
       \midrule 
    \multicolumn{6}{l}{\bf Approaches based on a latent survival times specification} \\
    \midrule      
   \multicolumn{1}{c}{\multirow{2}[0]{*}{Deep multi-task GPs}}& 
   \multicolumn{1}{c}{\multirow{2}[0]{*}{{Non-parametric}}}  & 
    \multicolumn{1}{c}{\multirow{2}[0]{*}{\ding{55}}} &  \multicolumn{1}{c}{\multirow{2}[0]{*}{\ding{51}}} &  
    \multicolumn{1}{c}{\multirow{2}[0]{*}{\ding{55}}} & \multicolumn{1}{c}{\multirow{2}[0]{*}{\ding{51}}}  \\   \multicolumn{1}{c}{\multirow{2}[0]{*}{}}  &         
    &  &       &     
    &       \\
          \multicolumn{1}{c}{\multirow{2}[0]{*}{Bayesian LDR}} & 
     \multicolumn{1}{c}{\multirow{2}[0]{*}{{Non-parametric}}} & 
    \multicolumn{1}{c}{\multirow{2}[0]{*}{\ding{55}}}& 
     \multicolumn{1}{c}{\multirow{2}[0]{*}{\ding{51}}} &
    \multicolumn{1}{c}{\multirow{2}[0]{*}{\ding{51}}} & \multicolumn{1}{c}{\multirow{2}[0]{*}{\ding{51}}} \\
    \multicolumn{1}{c}{\multirow{1}[0]{*}{}}   & & 
         &       &    
         &        \\

         \midrule 
    \multicolumn{6}{l}{\bf Others} \\
    \midrule      
      \multicolumn{1}{c}{\multirow{2}[0]{*}{Mixture models}}   &
    \multicolumn{1}{c}{\multirow{2}[0]{*}{Several}}
       &
    \multicolumn{1}{c}{\multirow{2}[0]{*}{\ding{55}}} &   
    \multicolumn{1}{c}{\multirow{2}[0]{*}{\ding{55}}} & 
    \multicolumn{1}{c}{\multirow{2}[0]{*}{\ding{55}}} &  \multicolumn{1}{c}{\multirow{2}[0]{*}{\ding{51}}}  \\   \multicolumn{1}{c}{\multirow{2}[0]{*}{}}  &         
    &  &       &      
    &       \\
\multicolumn{1}{c}{\multirow{1}[0]{*}{Tree-based Bayesian}}&
      \multicolumn{1}{c}{\multirow{2}[0]{*}{Semi-parametric}}& 
     \multicolumn{1}{c}{\multirow{2}[0]{*}{\ding{51}}} & \multicolumn{1}{c}{\multirow{2}[0]{*}{\ding{51}}} & 
     \multicolumn{1}{c}{\multirow{2}[0]{*}{ \ding{55}}} & \multicolumn{1}{c}{\multirow{2}[0]{*}{ \ding{51}}}  \\
      \multicolumn{1}{c}{\multirow{1}[1]{*}{mixture model}} &       &   
      &       &      
      &               \\
\multicolumn{1}{c}{\multirow{2}[0]{*}{Vertical modelling}}&
    \multicolumn{1}{c}{\multirow{2}[0]{*}{Semi-parametric}} &
    \multicolumn{1}{c}{\multirow{2}[0]{*}{\ding{55}}} &
    \multicolumn{1}{c}{\multirow{2}[0]{*}{\ding{51}}} &
    \multicolumn{1}{c}{\multirow{2}[0]{*}{\ding{51}}} & \multicolumn{1}{c}{\multirow{2}[0]{*}{\ding{51}}}   \\
     \multicolumn{1}{c}{\multirow{2}[0]{*}{}}  &         
    &  &       &     
    &       \\  
             \multicolumn{1}{c}{\multirow{2}[0]{*}{RSF}} &
       \multicolumn{1}{c}{\multirow{2}[0]{*}{Non-parametric}}&
     \multicolumn{1}{c}{\multirow{2}[1]{*}{\ding{51}}} & \multicolumn{1}{c}{\multirow{2}[1]{*}{\ding{51}}} & 
     \multicolumn{1}{c}{\multirow{2}[1]{*}{\ding{51}}} &
     \multicolumn{1}{c}{\multirow{2}[1]{*}{\ding{55}}}\\   \multicolumn{1}{c}{\multirow{2}[0]{*}{}}  &         
    &  &       &      
    &       \\


        \multicolumn{1}{c}{\multirow{2}[0]{*}{DSM}}  & 
     \multicolumn{1}{c}{\multirow{2}[0]{*}{Non-parametric\tnote{1}}}
    & 
    \multicolumn{1}{c}{\multirow{2}[0]{*}{\ding{51}}} &   \multicolumn{1}{c}{\multirow{2}[0]{*}{\ding{51}}} &  
    \multicolumn{1}{c}{\multirow{2}[0]{*}{\ding{55}}} & 
    \multicolumn{1}{c}{\multirow{2}[0]{*}{\ding{55}}} \\
  \multicolumn{1}{c}{\multirow{2}[0]{*}{}}  &         
    &  &       &      
    &       \\

             \midrule 
    \multicolumn{6}{l}{\bf CR survival models for discrete time-to-event data} \\
    \midrule     
         \multicolumn{1}{c}{\multirow{2}[0]{*}{BART}} & 
     \multicolumn{1}{c}{\multirow{2}[0]{*}{Non-parametric}} & 
     \multicolumn{1}{c}{\multirow{2}[0]{*}{\ding{51}}} & \multicolumn{1}{c}{\multirow{2}[0]{*}{\ding{51}}} & 
    \multicolumn{1}{c}{\multirow{2}[0]{*}{\ding{51}}} &  \multicolumn{1}{c}{\multirow{2}[0]{*}{\ding{55}}}     \\ 
    \\
   \bottomrule
    \end{tabular}}
    \begin{tablenotes}
     \item[1] A fully parametric specification is adopted for the survival model (mixture of Weibull or log-normal distributions), but a neural network learns a lower-dimensional representation for the covariates.
    \end{tablenotes}
      \label{tab:ML_summary}
      \end{threeparttable}
\end{table}
\endgroup


\section{Software and reproducibility} 
\label{S:Software}

Provision of open-source and well documented software is critical to ensure wide adoption of new statistical or machine learning methods. Towards this goal, \cite{MLR3PROBA} developed the 
\pkg{mlr3proba} \pkg{R} library, 
providing a common interface for several survival models, including some of the CR approaches here presented (removed from CRAN on May 2022, but actively maintained and available in GitHub). Another software resource was implemented by \cite{Mahani2019}, supporting Bayesian and non-Bayesian inference for cause-specific hazard models. 

Here, we summarise available software for the methods described in Section~\ref{S:CR} and~\ref{S:DiscreteCR}. While some of the approaches are 
available as \pkg{R} or \pkg{Python} packages, other methods are only accessible through \emph{ad hoc} source code in public repositories or, in the worse case scenario, there is no code available for the method's implementation. 
Table~\ref{tab:Software_summary} summarises this. In order to facilitate adoption, for the methods which have available R libraries, we provide vignettes to illustrate their usage using publicly available data \citep{pintilie2006competing}. Vignettes are available at \url{www.github.com/KarlaMonterrubioG/CompRisksVignettes}. 

Even when there are software packages accompanied with documentation and when analysis code is publicly available, reproducibility of an existing analysis is not guaranteed e.g.~due to differences in the computational environment \citep{beaulieu2017reproducibility}. Moreover, static vignettes or code included as part of a paper are not always updated as the associated software changes. This may introduce challenges when applying or benchmarking new methods. To ensure reproduciblity of the vignettes provided here, we also prepared a Docker image \citep{boettiger2015introduction} with all software requirements. The latter is available at: \url{https://github.com/KarlaMonterrubioG/CompRisksVignettes/pkgs/container/comprisksvignettes}.



 \begingroup
 \renewcommand{\arraystretch}{1.1}
\begin{table}[htbp]
\centering
\begin{threeparttable}
\caption{Software available for survival regression with CR.}
  \centering
  \small
   {\tabcolsep=4.25pt
   \begin{tabular}{lcllc}
    \toprule
    \multicolumn{1}{c}{\multirow{1}[1]{*}{Model}}  &   \multicolumn{1}{c}{\multirow{1}[1]{*}{CRAN}}    &
    \multicolumn{1}{c}{\multirow{1}[1]{*}{mlr3proba}}&
   \multicolumn{1}{c}{\multirow{1}[1]{*}{Other/comments}}\\
    \midrule
      \multicolumn{5}{l}{\bf Approaches based on a proportional cause-specific hazard specification} \\
    \midrule
          \multicolumn{1}{c}{\multirow{1}[1]{*}{Cox proportional}} &
       \multicolumn{1}{c}{\multirow{1}[1]{*}{\pkg{riskRegression}}}& 
      \multicolumn{1}{c}{\multirow{2}[1]{*}{\ding{51}}} &     \\
      \multicolumn{1}{c}{\multirow{1}[1]{*}{cause-specific hazard}} &    
       \multicolumn{1}{c}{\multirow{1}[1]{*}{ \pkg{survival}, \pkg{rms}}}&       &                \\
       
                 \multicolumn{1}{c}{\multirow{1}[1]{*}{Penalised Cox}} &
       \multicolumn{1}{c}{\multirow{2}[1]{*}{\pkg{glmnet}}}& 
      \multicolumn{1}{c}{\multirow{2}[1]{*}{\ding{51}}} &    \\
      \multicolumn{1}{c}{\multirow{1}[1]{*}{proportional hazard}} &    
     &       &                \\
              \multicolumn{1}{c}{\multirow{1}[1]{*}{Cox model-based}} &
       \multicolumn{1}{c}{\multirow{2}[1]{*}{\pkg{mboost}}}& 
      \multicolumn{1}{c}{\multirow{2}[1]{*}{\ding{51}}} &     \\
      \multicolumn{1}{c}{\multirow{1}[1]{*}{boosting}} &    &                   \\
          \multicolumn{1}{c}{\multirow{1}[1]{*}{Cox likelihood-based}} &
       \multicolumn{1}{c}{\multirow{2}[1]{*}{\pkg{Coxboost}}}& 
      \multicolumn{1}{c}{\multirow{2}[1]{*}{\ding{51}}} & Removed from CRAN\tnote{1}   \\ 
      \multicolumn{1}{c}{\multirow{1}[1]{*}{boosting}} &    &       &    \pkg{R} code available in GitHub\tnote{2}             \\
  
 \multicolumn{1}{c}{\multirow{2}[1]{*}{Lunn-McNeil}} &
       \multicolumn{1}{c}{\multirow{1}[1]{*}{\pkg{riskRegression}}}& 
      \multicolumn{1}{c}{\multirow{2}[1]{*}{\ding{51}}} &   \\
       &    
       \multicolumn{1}{c}{\multirow{1}[1]{*}{ \pkg{survival}, \pkg{rms}}}&       &   \\
     \midrule
       \multicolumn{5}{l}{\bf Approaches based on the CIF} \\
    \midrule  
     \multicolumn{1}{c}{\multirow{2}[1]{*}{Fine-Gray}} &
       \multicolumn{1}{c}{\multirow{1}[1]{*}{\pkg{riskRegression}}}& 
      \multicolumn{1}{c}{\multirow{2}[1]{*}{\ding{51}}} &     \\
       &    
       \multicolumn{1}{c}{\multirow{1}[1]{*}{ \pkg{cmprsk}}}&       \\
\multicolumn{1}{c}{\multirow{1}[1]{*}{Penalised proportional}} &
       \multicolumn{1}{c}{\multirow{2}[1]{*}{\ding{55}}}& 
      \multicolumn{1}{c}{\multirow{2}[1]{*}{\ding{55}}} &    \\
      \multicolumn{1}{c}{\multirow{1}[1]{*}{sub-distribution hazard}} &    
     &       &                \\
\multicolumn{1}{c}{\multirow{1}[1]{*}{Sub-distribution}} &
       \multicolumn{1}{c}{\multirow{2}[1]{*}{\pkg{Coxboost}}}& 
      \multicolumn{1}{c}{\multirow{2}[1]{*}{\ding{51}}} & Removed from CRAN\tnote{1}   \\ 
      \multicolumn{1}{c}{\multirow{1}[1]{*}{ hazard boosting}} &    &       &    \pkg{R} code available in GitHub\tnote{2}             \\

    \multicolumn{1}{c}{\multirow{1}[1]{*}{Pseudo-values}} &  \multicolumn{1}{c}{\multirow{1}[1]{*}{\pkg{pseudo}+\pkg{GEEPACK}}} &    
    \multicolumn{1}{c}{\multirow{1}[1]{*}{\ding{55}}}    & 
    \multicolumn{1}{l}{\multirow{1}[1]{*}{Implemented for $K=2$ only}}         \\  
    \multicolumn{1}{c}{\multirow{1}[0]{*}{Direct binomial}} &  \multicolumn{1}{c}{\multirow{1}[0]{*}{\pkg{timereg}}} & 
    \multicolumn{1}{c}{\multirow{1}[1]{*}{\ding{55}}} &
     \\

\multicolumn{1}{c}{\multirow{1}[0]{*}{Parametric constrained CIF}} &
    \multicolumn{1}{c}{\multirow{1}[0]{*}{\ding{55}}}& 
    \multicolumn{1}{c}{\multirow{1}[1]{*}{\ding{55}}}&          \multicolumn{1}{l}{\multirow{1}[0]{*}{{Example \pkg{R} code\tnote{3}}}}  \\
\multicolumn{1}{c}{\multirow{1}[0]{*}{Dependent DP}} &
    \multicolumn{1}{c}{\multirow{1}[0]{*}{\pkg{DPWeibull}}}& 
    \multicolumn{1}{c}{\multirow{1}[1]{*}{\ding{55}}}&  \multicolumn{1}{l}{\multirow{1}[1]{*}{Removed from CRAN\tnote{4}}}          \\
\multicolumn{1}{c}{\multirow{1}[0]{*}{SMTBoost}}  & 
     \multicolumn{1}{c}{\multirow{1}[1]{*}{\ding{55}}} & 
     \multicolumn{1}{c}{\multirow{1}[1]{*}{\ding{55}}}&       
          \\
\multicolumn{1}{c}{\multirow{1}[0]{*}{DeSurv}}  & 
     \multicolumn{1}{c}{\multirow{1}[1]{*}{\ding{55}}} & 
     \multicolumn{1}{c}{\multirow{1}[1]{*}{\ding{55}}}& 
       \multicolumn{1}{l}{\multirow{1}[1]{*}{ \pkg{Python} code available on  GitHub\tnote{5}}}
          \\
              \midrule
 \multicolumn{5}{l}{\bf Approaches based on a latent survival times specification} \\
    \midrule
\multicolumn{1}{c}{\multirow{1}[0]{*}{Deep multitask GPs}} &
     \multicolumn{1}{c}{\multirow{1}[0]{*}{\ding{55}}}& 
     \multicolumn{1}{c}{\multirow{1}[1]{*}{\ding{55}}}
     &    \\
\multicolumn{1}{c}{\multirow{1}[0]{*}{Bayesian LDR}} & 
  \multicolumn{1}{c}{\multirow{1}[1]{*}{\ding{55}}}& 
  \multicolumn{1}{c}{\multirow{1}[1]{*}{\ding{55}}}&
    \multicolumn{1}{l}{\multirow{1}[1]{*}{\pkg{R} code available in GitHub\tnote{6}}} & \\
         \midrule
       \multicolumn{5}{l}{\bf Others} \\
    \midrule       
  \multicolumn{1}{c}{\multirow{1}[0]{*}{Mixture models}} &     \multicolumn{1}{c}{\multirow{1}[0]{*}{\pkg{NPMLEcmprsk} }\tnote{7}} &    
  \multicolumn{1}{c}{\multirow{1}[1]{*}{\ding{55}}} &  
  \multicolumn{1}{l}{\multirow{1}[1]{*}{Example \pkg{R} code\tnote{8} ($K=2$) }} &    
   
  \\
\multicolumn{1}{c}{\multirow{1}[0]{*}{Tree-based Bayesian mixture model}}   & \multicolumn{1}{c}{\multirow{1}[1]{*}{\ding{55}}}
     &  \multicolumn{1}{c}{\multirow{1}[1]{*}{\ding{55}}}&
     \multicolumn{1}{l}{\multirow{1}[1]{*}{Example \pkg{R} code\tnote{9} ($K=2$) }} \\
 \multicolumn{1}{c}{\multirow{1}[0]{*}{Vertical modelling}}&
    \multicolumn{1}{c}{\multirow{1}[0]{*}{\pkg{splines}+\pkg{survival}}} & \multicolumn{1}{c}{\multirow{1}[1]{*}{\ding{55}}}
        &     \multicolumn{1}{l}{\multirow{1}[1]{*}{Example \pkg{R} code\tnote{8} }}   \\
    \multicolumn{1}{c}{\multirow{1}[0]{*}{RSF}} &
    \multicolumn{1}{c}{\multirow{1}[0]{*}{\pkg{randomForestSRC}}}&  \multicolumn{1}{c}{\multirow{1}[1]{*}{\ding{51}}}&          \\
  
   \multicolumn{1}{c}{\multirow{1}[0]{*}{DSM}}  & \multicolumn{1}{c}{\multirow{1}[0]{*}{\ding{55}}} &
   \multicolumn{1}{c}{\multirow{1}[0]{*}{\ding{55}}} &
   \multicolumn{1}{l}{\multirow{1}[0]{*}{\pkg{Python} package \pkg{dsm}\tnote{10}}}\\
        \midrule
       \multicolumn{5}{l}{\bf  CR survival models for discrete time-to-event data} \\
    \midrule       
\multicolumn{1}{c}{\multirow{1}[0]{*}{BART}} &  \multicolumn{1}{c}{\multirow{1}[0]{*}{\pkg{BART}}} &
    \multicolumn{1}{c}{\multirow{1}[1]{*}{\ding{55}}}&
    \\
     \bottomrule
    \end{tabular}
\begin{tablenotes}
 \item[1] \url{https://cran.r-project.org/web/packages/CoxBoost/index.html}
 \item[2] \url{https://github.com/binderh/CoxBoost}.  
 \item[3] See supplementary material in \cite{shi2013} 
   \item[4] \url{https://cran.r-project.org/web/packages/DPWeibull/index.html}
  \item[5] \url{https://github.com/djdanks/DeSurv}  
 \item[6] \url{https://github.com/zhangquan-ut/Lomax-delegate-racing-for-survival-analysis-with-competing-risks}
  \item[7] This library implements the approach by \cite{Chang2007Nonparametric}. It is available in CRAN (last updated in 2018) but the provided example does not run. We have not contacted the authors to resolve this issue. 
 \item[8] See supplementary material in \cite{haller2013}
 \item[9] \url{https://github.com/alexisbellot/HBM}
 \item[10] \url{https://autonlab.github.io/DeepSurvivalMachines} 

\end{tablenotes}
}
  \label{tab:Software_summary}%
\end{threeparttable}
\end{table}%
\endgroup


   

\section{Evaluating performance} \label{S:benchmark}
When proposing a new method, researchers are often interested in evaluating and comparing its performance. For example, for (semi-)parametric models, one may use synthetic data to assess whether parameter estimates are unbiased. For approaches that include variable selection, one may evaluate their ability to identify a correct set of input variables. When the goal is to perform risk prediction, the emphasis is on evaluating how well a method is able to predict {\it whether} and/or {\it when} specific event types will occur. To evaluate predictive performance, an external (or test) dataset that was not used to fit the model could be used. However, internal validation (e.g.~via boostrapping or cross-validation) is also important, particularly for small datasets or when the number of observed events is small \citep{steyerberg2016prediction}.

Recently, \cite{vanGelovene2022} discussed how to evaluate predictive performance in competing risks settings, providing examples in \pkg{R} 
(see \url{https://github.com/survival-lumc/ValidationCompRisks}). 
They focused on cases in which the goal is to predict {\it whether} the event of interest will occur within a given time-frame (e.g.~5-year survival). \cite{vanGelovene2022} emphasised the need to evaluate different aspects of predictive performance including {\it{calibration}}, something that is often overlooked when developing risk prediction models \citep{van2019calibration}. 
A well calibrated model will assign the correct event probability at all levels of predicted risk. 
Another important aspects are {\it discrimination}, i.e.~whether the model assigns a higher risk to individuals who experience the event earlier. 
Here, we briefly describe some of the metrics that can be used to evaluate these aspects. 


\paragraph{Concordance.} A popular metric to assess discrimination in the context of survival models is via a \emph{concordance index} \cite[also referred to as C-index,][]{harrell1982}. Generally, higher C-index indicates better discrimination (and a value equal to 0.5 indicates no discrimination ability). Several definitions are available, including some that have been adapted to CR settings. For example, if the aim is to predict whether $k$th event type is observed prior to a pre-specified time $\tau$, \cite{wolbers2014} proposed the following cause-specific time-dependent C-index:
\begin{equation} \label{eq:Cstatistic}
    C_k(\tau) = \Pr \left( \CIF_k( \tau \mid \bx_i) >   \CIF_k(\tau \mid \bx_j)\mid \{ Z_i=k\} \land \{ T_i \leq \tau\} \land \{ T_i \leq T_j \lor Z_j \neq k \} \right),
\end{equation} for a random pair of individuals ($i$ and $j$). This metric quantifies if model is able to correctly rank the risk of observing the . 
More recently, \citet{Mihaela2019} proposed a joint concordance index to evaluate the model's ability to correctly predict both the event type and time. Their approach may be of interest in cases where more than one event type is of interest. If the interest is to assess discrimination across the whole follow-up period rather than at a specific time-point $\tau$, a weighted average of $C_k(\tau)$ could be used (see e.g.~the approach proposed by \cite{antolini2005} for a single event type).

\paragraph{Brier score.} 
    \cite{Schoop2011} adapted the proper scoring score introduced by \cite{graf1999} to competing risks settings. For a given prediction time $\tau$, the Brier score for cause $k$ is defined as the 
    a weighted average of the squared differences between the cause-specific event indicators and the predicted cause-specific survival probabilities: \begin{equation}
     \text{BS}_k(\tau) = \tfrac{1}{n} \sum_{i=1}^{n} w_i  \left[ \mathbbm{1}{ \{T_i \leq \tau, Z_i= k \}} - \Pr(T_i \leq \tau, Z_i=k \mid \bx_i) \right]^2, \label{eq:BS}  
    \end{equation}
    where the weights $w_i$ are used to account for right censoring \citep[][, Theorem 4.1]{Schoop2011}. 
This can be interpreted as a metric of overall performance, as it encompasses both calibration and discrimination. 
   The lower the value of \eqref{eq:BS}, the better.
    To summarise performance across a range of time-points, an integrated Brier Score can be defined \citep{graf1999}.
    
    The absolute value of \eqref{eq:BS} is difficult to interpret as its scale depends on the number of observed events. As an alternative, 
    an scaled version of \eqref{eq:BS} can be used. 
    The scaled Brier score can be computed as follows \citep{vanGelovene2022}: 
\begin{equation}
    \text{BS}_k(\tau)^{\text{scaled}} = 1 - \frac{\text{BS}_k(\tau)}{\text{BS}_k(\tau)^{\text{null}}},
\end{equation}
where ${\text{BS}_k(\tau)^{\text{null}}}$ denotes the Brier score under the null model (no covariates) and which can be computed using the Aalen-Johansen estimator \citep{AJ1978}. The later lies between 0 and 1, where 1 indicates perfect predictions. 

\section{Discussion}
 
We summarised a broad range of competing risks modelling techniques that encompass traditional and state-of-the-art approaches.
Our objective is to provide the reader with a synthesised catalogue, with unified notation and interpretation. We also briefly review metrics that can be used to evaluate and compare predictive performance. 
We emphasise that when deciding on the appropriate CR method to employ, the practitioner
needs to carefully consider the specific research question at hand, as there is no single approach that works well for all applications. For instance, \citet{Austin2016} highlight that a CIF formulation is more appropriate for prognosis models; whereas,
a CS approach is better suited 
to resolve etiological questions. Moreover, in some cases, reporting both methods can provide useful insights of the covariate effects on, both, the incidence and the rate of occurrence of the event \citep{LATOUCHE2013648}. 


 
In order to promote the usage of state-of-the-art approaches, we point out to available software and, demonstrate its practical implementation through reproducible \pkg{R} vignettes.
Emphasis on reproducibility is critical when developing and evaluating new methods.
While
making the implementation of the method publicly available using version control hosting tools; such as GitHub or BitBucket, helps towards this goal;
this is not enough.
The code must be well documented and, when possible, accompanied with the raw data (synthetic or real) that was used to assess performance. It is also important provide details on how such data was generated or processed, as well as a clear description of any {\it ad hoc} choices made (e.g.~inclusion/exclusion criteria). 
For instance, 
the Surveillance, Epidemiology, and End Results (SEER) Program\footnote{https://seer.cancer.gov} datasets have been employed to showcase several CR methods
\citep[e.g.~][]{zhang2018nonparametric, Alaa2017deep, TMixtureM2018, Siames2018, Bellot2018}. However, detailed information on how the dataset used was 
preprocessed is usually not provided (in some cases, authors do not
even provide the full list of covariates used in the analysis). Similar issues have been reported when using the MIMIC database \citep{pmlr-v68-johnson17a}. More systematic and reproducible benchmark pipelines \citep{mangul2019systematic} for competing risks methods are urgently required to reduce the gap between developers and users. 


\section*{Acknowledgments}
KMG was supported by an MRC University Unit grant to the MRC Human Genetics Unit. NC-C was supported by the Medical Research Council and University of Edinburgh via a Precision Medicine PhD studentship (MR/N013166/1). CAV was supported by a Chancellor's Fellowship provided by The University of Edinburgh. CAV was also supported by a British Heart Foundation-Turing Cardiovascular Data Science Award (BCDSA/100003).
For the purpose of open access, the author has applied a CC-BY public copyright licence to any Author Accepted Manuscript version arising from this submission.
The authors would like to acknowledge the support of Rodney Sparapani, Shu-Kay Angus Ng and Geoffrey McLachlan. They kindly provided insight about their methods, and shared and/or pointed out to code for their implementation. 

\newpage

\bibliographystyle{unsrtnat}
\bibliography{references}  

\newpage






\section*{Supplementary material to A review on competing risks methods for survival analysis}

\subsection{Appendix - Parameter estimation} \label{Sup:A}

For the $k$-th event type, the corresponding regression coefficients, $\bbeta_k$, in \eqref{eq:CPH} can be estimated by maximising the partial likelihood:
\begin{equation} 
    \mathcal{L}^{\text{CS}}(\bbeta_k) = \prod_{i=1}^n \left( \frac{ \exp( \bx^\top_{i} \bbeta_{k} )}{ \sum_{l \in R_i} \exp(\bx^\top_{l} \bbeta_{k}) } \right)^{ \mathbbm{1}\{Z_i = k \}},
    \label{eq:lik_CS}
\end{equation}
where $R_i$ denotes the set of observations at risk at time $t_i$, i.e. subjects that are not censored or that have not experienced a competing event by time $t_i$.

In contrast, the regression parameters in \eqref{eq:FG}, $\bm{\gamma}_k$, are obtained by maximisation of the pseudo-likelihood function:
 \begin{equation} 
     \mathcal{L}^{\text{FG}}(\bm{\gamma}_k) = \prod_{i=1}^n \left( \frac{ \exp( \bx^\top_{i} \bm{\gamma}_{k} )}{ \sum_{l \in \tilde{R}_i} w_{il}\exp(\bx^\top_{l} \bm{\gamma}_{k})} \right)^{  \mathbbm{1}\{Z_i = k \} }, \label{eq:lik_FG}
     \end{equation}
where $\tilde{R}_i$ denotes the set of observations at risk at time $t_i$, i.e. subjects that are event free as well as subjects that have already experienced a competing event by time $t_i$. In addition, $ w_{il}$ denotes subject-specific weights which are set to one for those individuals that do not experience an event of interest and are given by
$$w_{il}= \frac{S^{\text{KM}}(t_i)}{ S^{\text{KM}}( \min(t_l,t_i))},$$ for subjects experiencing a competing event, where $S^{\text{KM}}(\cdot)$ denotes the  Kaplan-Meier estimate of the survival function.


\subsection{Appendix - Cox boosting}\label{Sup:B}
Cox model-based boosting ({\it{mboost}}) and Cox likelihood based boosting ({\it{Coxboost}}) differ on how the corresponding regression coefficients are estimated at every iteration, $b=1, \ldots, B.$
Specifically, {\it{mboost}} utilise least squares estimators, $\hat{\alpha}_j ^{(b)}$ ($j=1,\ldots,p$), as weak learners; such that, the updating step corresponds to $\beta^{(b)}_{kj^*} =  \beta^{(b-1)}_{kj^*} + \lambda \hat{\alpha}^{(b)}_{j^*}$, with $j^*$ denoting the best update selected by minimisation of the residual sum of squares, and $0<\lambda\leq 1$ the step-length or learning rate \citep[see Section 5 of][ for an empirical study to select $\lambda$]{ friedman2001}.
Instead,{\it{Coxboost}} employs the negative gradient to obtain the score function and the observed Fisher information which in turn are used to derive the weak learners, $\hat{\eta}_j ^{(b)}$, that will be used in the updates $\beta^{(b)}_{kj^*} = \beta^{(b-1)}_{kj^*} + \hat{\eta}^{(b)}_{j^*}$, where the best update $j^*$ is that with the largest reduction for the penalised partial log-likelihood $\ell^{\text{pen}}( \beta^{(b)}_{j}) = \ell^{\text{CS}}( \beta^{(b)}_{j}) + \frac{\lambda}{2} {\beta_{j}^{(b)}}^{2}.$ Note that in this case the penalisation is directly included in $\hat{\eta}_j ^{(b)}$.

\subsection{Appendix - Lunn-McNeil augmented layout}
\label{Sup:C}
Assume we have two event types, $K=2$. The following table shows the observed data for 3 subjects. The first, experienced event type $2$ at time $10$, the second is assumed to be censored by time $70$, and the third experienced event type $1$ at time $14$.
\begin{table}[htbp]
\caption{Original layout}
  \centering
  \small
   {\tabcolsep=4.25pt
   \begin{tabular}{ccccc}
    \toprule
    Individual & Event time ($T$)  &  Event type ($Z$)  &  Covariates  \\
    \midrule
     $1$ & $10$ & 2   &  $\bx^\top_1$\\
    $2$ & $70$ & 0   & $\bx^\top_{2}$ \\
     $3$ & $14$ & 1   & $\bx^\top_{3}$ \\
     \bottomrule
    \end{tabular}}
\end{table}


The augmented layout required for LM approach necessitates to have $2$ rows per subject, one for each competing event. In addition, we add event type indicators $\delta_{ik}$.
\begin{table}[htbp]
\caption{Augmented layout for LM}
  \centering
  \small
   {\tabcolsep=4.25pt
   \begin{tabular}{cccccc}
    \toprule
    \multicolumn{1}{c}{\multirow{2}[1]{*}{Individual}} &  \multicolumn{1}{c}{\multirow{2}[1]{*}{Event time ($T$)}} & 
   \multicolumn{1}{c}{\multirow{2}[1]{*}{Event type ($Z$)}} &
    \multicolumn{2}{c}{{Event type indicator ($\delta$)}} &
     \multicolumn{1}{c}{\multirow{2}[1]{*}{Covariates}} \\
    & & &
    \multicolumn{1}{c}{\multirow{1}[1]{*}{$Z = 1$}}  &   
    \multicolumn{1}{c}{\multirow{1}[1]{*}{$Z = 2$}} \\
    \midrule
    $1$ & $10$ & 2 &0& 1& $\bx^\top_1$ \\
    $ 1$ & $10$ & 0 & 1& 0 & $\bx^\top_1$ \\ \hdashline
    $2$ & $70$ & 0  & 1 &0 &$\bx^\top_{2}$\\
    $ 2$ & $70$ & 0 & 0 & 1&$\bx^\top_{2}$\\
     \hdashline
    $3$ & $14$ & 1  & 1 &0 & $\bx^\top_{3}$\\
    $3$  & $14$ & 0 & 0 & 1&$\bx^\top_{3}$\\
     \bottomrule
    \end{tabular}}
\end{table}


\subsection{Appendix - R vignettes}
Vignettes showcasing the usage of some methods
are available online at: 
\url{https://github.com/KarlaMonterrubioG/CompRisksVignettes}.

\end{document}